\RequirePackage{amsmath}
\documentclass[runningheads]{llncs}
\usepackage{microtype}
\usepackage{graphicx}
\usepackage{subfigure}
\usepackage{booktabs} 
\usepackage{tikz}
\usetikzlibrary{arrows,positioning}
\usepackage{amsfonts}
\usepackage{IEEEtrantools}  
\tikzset{
	terminal/.style={
		minimum size=6mm,
		font=\ttfamily}
}

\usepackage{todonotes}

\usepackage{booktabs}

\usepackage{hyperref}

\newtheorem{defn}{Definition}

\newcommand{\E}{\mathbb{E}}
\newcommand{\Prob}{\mathbb{P}}

\newcommand{\Cov}{\text{Cov}}

\newcommand{\I}{\mathbf{I}}

\newcommand{\Hm}{\mathbf{H}}

\newcommand{\X}{\mathbf{X}}
\newcommand{\x}{\mathbf{x}}
\newcommand{\Y}{\mathbf{Y}}

\newcommand{\W}{\mathbf{W}}
\newcommand{\w}{\mathbf{w}}

\newcommand{\bS}{\mathbf{S}}
\newcommand{\s}{\mathbf{s}}

\newcommand{\multicol}[2]{\IEEEeqnarraymulticol{#1}{#2}}

\begin{document}
	
	\title{Impartial Predictive Modeling and the Use of Proxy Variables}

\author{Kory D. Johnson\inst{1}\orcidID{0000-0002-7322-2451} \and
Dean P. Foster\inst{2} \and
Robert A. Stine\inst{3}}
\authorrunning{K. Johnson et al.}
\institute{TU Wien, 1040 Vienna, Austria\\
\email{kory.johnson@tuwien.ac.at} \and
Amazon, New York, New York, USA\\ \email{dean.foster@gmail.com} \and
University of Pennsylvania, Philadelphia, Pennsylvania, USA\\
\email{stine@wharton.upenn.edu}}

\begin{abstract}
Fairness aware data mining (FADM) aims to prevent algorithms from 
discriminating 
against protected groups. The literature has come to an impasse as to what 
constitutes explainable variability as opposed to discrimination. This 
distinction hinges on a rigorous understanding of the role of proxy variables; 
i.e., those variables which are associated both the protected feature and the 
outcome of interest. We demonstrate that fairness is achieved by ensuring 
\emph{impartiality} with respect to sensitive characteristics and provide a 
framework for impartiality by accounting for different perspectives on the data 
generating process. In particular, fairness can only be precisely defined in a 
full-data scenario in which all covariates are observed. We then analyze how 
these models may be conservatively estimated via regression in partial-data 
settings. Decomposing the regression estimates provides insights into 
previously 
unexplored distinctions between explainable variability and discrimination that 
illuminate the use of proxy variables in fairness aware data 
mining.
\end{abstract}

\maketitle

\section{Introduction}

Machine learning has been a boon for improved decision making. The increased 
volume and variety of data has led to a host of data mining tools 
for knowledge discovery; however, automated decision making using vast 
quantities of data needs to be tempered by caution. One goal in FADM is to 
provide suitably ``fair" estimates of a response $Y$, given legitimate 
covariates $\x$, sensitive covariates $\s$, and suspect covariates $\w$. The 
primary distinctions between these categories concerns the ability of an 
individual to be morally responsible for their value: covariates in 
$\s$ are considered to be outside of one's control, e.g. race and gender, while 
$\x$ are those features for which an individual can be held accountable. The 
$\w$ group contains those covariates for which it is uncertain whether or 
not one ought to be held responsible for their value.

These covariate groups and the response $Y$ are connected through an unknown, 
joint probability distribution $\Prob(Y,\x,\s,\w)$, and our data consists of 
$n$ iid draws from this joint distribution. The standard statistical goal is to 
estimate the conditional expectation of $Y$ given the covariates:
\begin{IEEEeqnarray*}{rCl}
	Y & = & \E[Y|\x,\s,\w] + u
\end{IEEEeqnarray*}
where $u$ has mean zero and is uncorrelated with all functions of the
covariates. 

Our goal is to estimate a conditional expectation that is ``fair'' with respect 
to the sensitive covariates. Others have argued for penalizing discrimination 
during estimation \cite{CalV10,Cald+13,Kam+12b}, modifying the input data 
before supervised training takes place
\cite{PedRT08,HajD13,BenH19}, or modifying objective functions with fairness 
criteria \cite{KamCP10,Zem+13,Heidari+18,Heidari+19}. Conceptually closest to 
our work is \cite{Heidari+19}, which explicitly discusses economic models of 
equality of opportunity and presents an optimization problem of maximizing
utility subject to a constraint on prediction error. Their formulation 
incorporates the concept of ``effort-based'' utility to encode the effects of 
legitimate covariates. 

From a technical perspective, \cite{Kilbertus17,Kusner17,Chiappa2018,Ravi+20} 
use similar path diagrams as those presented in Section \ref{sec:models} to 
describe fair estimation methods. Our focus, however, is different, as we 
concretely describe heretofore unknown distinctions between covariate groups. 
This understanding is crucial to correctly determining these groups, which are 
considered to be externally given in previous analyses. Other papers which 
address this issue include \cite{PopeS11,Adler18}, though the present paper 
covers more settings (none, some, or all suspect covariates), broader use cases 
(extending to ``black-box'' models), and defends the disentanglement and partial 
omission of the proxy signal.

A simple example clarifies the issue of fairness. Consider a bank that wants to 
estimate the risk of a loan applicant via a credit score model. Such automated 
eligibility or pricing systems are ubiquitous both in banking and public 
assistance offices \cite{Eubanks18}. The concern is  that personalized credit 
pricing could provide results which either reflect discrimination 
which is present in the training data or either intentionally or unintentionally 
discriminate based on race or gender etc. Fair lending law attempts to address 
this through ``input scrutiny'', in which ``sensitive'' or ``protected'' 
covariates such as race, gender, and age are barred from use \cite{Gillis21}. 
The goal of input scrutiny is to model prices solely on the remaining 
``legitimate'' covariates, i.e., the log of historical credit use. The use of 
big data in pricing models, however, has opened the door to non-standard data 
sources such as purchase history and online activities, which could allow for 
predatory or targeted pricing \cite{HurleyA16}. This highlights a third 
covariate group of ``suspect'' or ``potentially discriminatory'' covariates 
whose information content for establishing creditworthiness is uncertain and 
potentially serves only to establish ``creditworthiness by  association,'' i.e., 
by associating an applicant to a protected covariate \cite{HurleyA16}. The 
canonical example of such a proxy variable is the applicant's address. While 
location is not a protected characteristic such as race, it is often barred from 
use given the ability to discriminate using it. 


FADM asks whether or not the estimates the bank constructs are fair and 
perhaps even what effect enforcing fairness has on profit \cite{KozJL21}. This 
is different than asking if the data are fair or if the historical practice of 
giving loans was fair. It is a question pertaining to the estimates produced by 
the bank's model, and thus necessarily would imply a shift in lending law to 
outcome-focused analysis \cite{Gillis21}. This generates several questions. 
First, what does fairness mean for this statistical model? Second, what should 
the role of the sensitive covariates be in this estimate? Third, how do 
legitimate and suspect covariates differ? Lastly, how do we constrain the use of 
the sensitive covariates in black-box algorithms? These questions are addressed 
in the remainder of the paper.
 
A primary hurdle for FADM is to separate explainable variability from 
discrimination. Many authors have argued that input scrutiny 
is insufficient to achieve fairness \cite{Kam+12b,KamZC13,Gillis21}. Due to the 
relationships between race and other covariates, merely removing race can leave 
lingering discriminatory effects that permeate the data and potentially 
perpetuate discrimination. The use of covariates associated with $\s$ to 
discriminate is called ``redlining'', which originated in the United States to 
describe maps that were color-coded to represent areas in which banks would not 
invest. While policies were facially neutral and race blind, they were in-effect 
discriminatory. The advent of big data has given rise to what has been called 
technological or digital redlining \cite{Noble18}, wherein similar demarcations 
can be made by, for example, only looking for housing close to high-quality 
schools. Conceptually, the core issue is the \emph{misuse} of available 
information. 

We improve upon previous discussions by providing a 
simple, tractable formulation of impartiality that addresses issues often 
encountered in real data; namely, that sensitive features often \emph{do not 
appear} to be unrelated to the response. In what follows, we argue that this 
task is accomplished by creating impartial estimates. Intuitively, impartiality 
requires that the sensitive covariates do not influence estimates. For clarity, 
we will refer to this as the fairness assumption:

\paragraph{Fairness Assumption:} Sensitive covariates ought not be a relevant 
source of variability or merit.
\\

A clear objection to this assumption is that it is often not observed in the 
data, but this is precisely the point. To be explicit, consider a circumstance 
in which $Y$ is an observed quantity realized by an agent in a ``free'' 
(non-coerced) way, e.g. credit history. To play devil's advocate in this case, 
a 
most fundamental question for FADM is: if models are intended to describe the 
world, and the fairness assumption is often inaccurate, why compromise 
predictive accuracy for fairness' sake? Instead of addressing this 
philosophically, we specify models in which fairness is the accurate 
statistical description of the world. Our fairness assumption is specified as 
an ``ought'' statement, as it embodies this often unrealized ideal.

Impartial estimates are fair because the covariate groups are chosen to be 
normatively relevant and are assumed to be provided externally. As such, the 
statistical task is disjoint from the normative task of identifying covariate 
groups. 
This project uses the term ``impartial'' to describe the statistical 
goal. This is done in order to separate our task from normative complications.  
That being said, the different covariate groups have normative significance and 
need to be differentiated. 

The construction of our impartial estimates is motivated by distinctions in the 
philosophical literature on equality of opportunity, which analyzes the way in 
which benefits are allocated in society \cite{sep-eo}. One way of understanding 
equality of opportunity is formal equality of opportunity (FEO), which requires 
an open-application for benefits (anyone can apply) and that benefits are 
awarded to those of highest merit. Merit will of course be measured differently 
depending on the benefit in question. There may, however, be cause for concern 
if discrimination exists in the analysis of merit or the ability of some 
individuals to be of high merit.

Substantive equality of opportunity (SEO) contains the same strictures as 
above, 
but is satisfied only if everyone has a genuine opportunity to be of high 
merit. 
In particular, suppose there are social restrictions or benefits that only 
allow 
one group to be of high merit. In this case, proponents of SEO claim that true 
equality of opportunity has not been achieved. While many countries lack a 
formal system to enforce this, some may argue that cycles of poverty and wealth 
lead to a similar regress in the reasons for the disparity between protected 
groups.

Identifying impartial estimates and fairness constructively
has both philosophical and empirical support. The philosopher John Rawls
discusses fair institutions as the method of achieving fairness
\cite{sep-rawls}. Similarly, \cite{Brock06,Grgic+16} explain the importance of 
process fairness as opposed to merely outcome fairness. Process fairness focuses 
on how people are treated throughout the process of a decision. The authors 
identify several examples in which firms attempt to layoff workers in a fair 
manner. Workers often feel that the decisions are fair when they are consulted 
frequently and the process is transparent, even if their severance packages are 
far worse. This points out the importance of fair treatment as fair use of 
information, not merely a measure of the outcome.

Our main contribution is a framework which allows for rigorous analysis of how 
the use of covariates changes when moving between FEO and SEO. We address the 
complications produced by the various viewpoints by introducing and 
demonstrating the importance of the suspect covariate group $\w$. After 
providing this framework, it will be clear how to both construct impartial 
estimates using simple procedures as well as correct black-box estimates. 

Section \ref{sec:models} introduces various data generating models in order to 
construct mathematical models of impartiality. This section also draws 
connections to the literature on causal inference. Importantly, the claim that 
these are impartial requires a strict interpretation of the influence of 
sensitive covariates. Section \ref{sec:create} provides a simple procedure for 
constructing impartial estimates.  The difference between achieving group 
fairness via FEO and SEO is demonstrated via a simple example in Section 
\ref{sec:simple-ex}. Section \ref{sec:data} analyzes two data sets from 
criminology that consider the effect of race and sex. Impartial estimates are 
produced with an package R \cite{R_2019} that is available on github.com.

\section{Mathematical Models of  Impartiality}
\label{sec:models}

Fairness in modeling will be explained via path diagrams to conveniently 
represent conditional independence assumptions. While often used as a model to 
measure causal effects, we are explicitly not using them for this purpose. This 
stems from a different object of interest: in causal modeling, one cares about 
a 
casual parameter or direct effect of the covariate of interest whereas we care 
about the estimates produced by the model. Estimating a causal effect requires 
considering a counterfactual, such as a patient's outcome under the treatment 
even though they were part of the control group. See, for example, 
\cite{morgan_winship_14}.

Our goal is to create impartial \emph{estimates}, whereas the estimation of a 
causal effect would attempt to answer whether the historical data are 
impartial. 
Hence, we do not require the same type of causal interpretation, which is 
fraught with difficulties for attributes such as race \cite{Holl03,BenH19}. 
Counterfactuals can be easily computed because it only requires 
producing an estimate for a modified observation, regardless of whether it 
exists in the data set. We do 
not need recourse to the interventionist or causal components of standard 
causal 
models \cite{Pearl09} and can focus only on their predictive component. Path 
diagrams only represent the conditional independence assumptions made between 
variables as necessitated by the fairness assumption.

The rest of this section briefly introduces impartial estimates in stages via 
models in which the fairness assumption is tractable. We begin by enforcing 
FEO, 
which only uses sensitive and legitimate covariates. The goal in FEO is to have 
a best estimate of merit while satisfying the legal requirements of disparate 
treatment and disparate impact. Second, we consider a full SEO model, in which 
there are no legitimate covariates, only sensitive and suspect covariates. The 
total model case with sensitive, legitimate, and suspect covariates is 
considered last.

\begin{table}
	\caption{Observationally Equivalent Data Generating Models: FEO (first 
row), 
SEO (second row), mixture (third row).}
	\label{tab:diagrams}
  \begin{center}
    \begin{small}
      \begin{sc}
\begin{tabular}{c|c|c}
  \toprule
	Observed & Unrestricted & Fair \\ 
	\midrule
	\begin{tikzpicture}[node distance = 2mm and 2mm]
	\node (s) [terminal]                                  {$\s$};
	\node (xo) [terminal, above right=of s]               {$\x_o$};
	\node[transparent] (xu) [terminal, below right=of s]            
	{$\x_u$};
	\node (d) [terminal, below right=of xo]               {$Y$};
	\path (s) edge[->] (xo) edge[->] (d);
	\path (xo) edge[->] (d);
	\end{tikzpicture} &
	\begin{tikzpicture}[node distance = 2mm and 2mm]
	\node (s) [terminal]                                  {$\s$};
	\node (xo) [terminal, above right=of s]               {$\x_o$};
	\node (xu) [terminal, below right=of s]               {$\x_u$};
	\node (d) [terminal, below right=of xo]               {$Y$};
	\path (s) edge[->] (xo) edge[->] (xu) edge[->] (d);
	\path (xo) edge[->] (d) edge[<->,dashed] (xu);
	\path (xu) edge[->] (d);
	\end{tikzpicture} &
	\begin{tikzpicture}[node distance = 2mm and 2mm]
	\node (s) [terminal]                                  {$\s$};
	\node (xo) [terminal, above right=of s]               {$\x_o$};
	\node (xu) [terminal, below right=of s]               {$\x_u$};
	\node (d) [terminal, below right=of xo]               {$Y$};
	\path (s) edge[->] (xo) edge[->] (xu);
	\path (xo) edge[->] (d) edge[<->,dashed] (xu);
	\path (xu) edge[->] (d);
	\end{tikzpicture}\\
	\midrule
	\begin{tikzpicture}[node distance = 2mm and 2mm]
	\node (s) [terminal]                                  {$\s$};
	\node (w) [terminal, below=of s]                      {$\w$};
	\node (d) [terminal, above right=of xu]               {$Y$};
	\path (s) edge[->] (w) edge[->] (d);
	\path (w) edge[->] (d);
	\end{tikzpicture} &
	\begin{tikzpicture}[node distance = 2mm and 2mm]
	\node (s) [terminal]                                  {$\s$};
	\node (w) [terminal, below=of s]                      {$\w$};
	\node (xu) [terminal, below right=of s]               {$\x_u$};
	\node (d) [terminal, above right=of xu]               {$Y$};
	\path (s) edge[->] (xu) edge[->] (w) edge[->] (d);
	\path (w) edge[<->,dashed] (xu) edge[->] (d);
	\path (xu) edge[->] (d);
	\end{tikzpicture} &
	\begin{tikzpicture}[node distance = 2mm and 2mm]
	\node (s) [terminal]                                  {$\s$};
	\node (w) [terminal, below=of s]                      {$\w$};
	\node (xu) [terminal, below right=of s]               {$\x_u$};
	\node (d) [terminal, above right=of xu]               {$Y$};
	\path (s) edge[->] (xu) edge[->] (w);
	\path (w) edge[<->,dashed] (xu);
	\path (xu) edge[->] (d);
	\end{tikzpicture}\\
	\midrule
	\begin{tikzpicture}[node distance = 2mm and 2mm]
	\node (s) [terminal]                                  {$\s$};
	\node (w) [terminal, below=of s]                      {$\w$};
	\node (xo) [terminal, above right=of s]               {$\x_o$};
	\node (d) [terminal, below right=of xo]               {$Y$};
	\path (s) edge[->] (xo) edge[->] (d) edge[->] (w);
	\path (w) edge[<->,dashed] (xo) edge[->] (d);
	\path (xo) edge[->] (d);
	\end{tikzpicture} &
	\begin{tikzpicture}[node distance = 2mm and 2mm]
	\node (s) [terminal]                                  {$\s$};
	\node (w) [terminal, below=of s]                      {$\w$};
	\node (xo) [terminal, above right=of s]               {$\x_o$};
	\node (xu) [terminal, below right=of s]               {$\x_u$};
	\node (d) [terminal, above right=of xu]               {$Y$};
	\path (s) edge[->] (xo) edge[->] (xu) edge[->] (w) edge[->] (d);
	\path (w) edge[<->,dashed] (xo) edge[<->,dashed] (xu) edge[->] (d);
	\path (xo) edge[->] (d) edge[<->,dashed] (xu);
	\path (xu) edge[->] (d);
	\end{tikzpicture} &
	\begin{tikzpicture}[node distance = 2mm and 2mm]
	\node (s) [terminal]                                  {$\s$};
	\node (w) [terminal, below=of s]                      {$\w$};
	\node (xo) [terminal, above right=of s]               {$\x_o$};
	\node (xu) [terminal, below right=of s]               {$\x_u$};
	\node (d) [terminal, above right=of xu]               {$Y$};
	\path (s) edge[->] (xo) edge[->] (xu) edge[->] (w);
	\path (w) edge[<->,dashed] (xo) edge[<->,dashed] (xu);
	\path (xo) edge[->] (d) edge[<->,dashed] (xu);
	\path (xu) edge[->] (d);
	\end{tikzpicture}
	\\\bottomrule
\end{tabular} 
\end{sc}
\end{small}
\end{center}
\vskip -0.1in
\end{table}

FEO is not concerned with potentially discriminatory covariates. 
Consider an idealized population model that includes all possible covariates: 
$\x_{o}$ contains the observed, legitimate covariates, and $\x_{u}$ contains 
the unobserved, legitimate
covariates. Unobserved covariates could be potentially
observable such as drug use, or unknowable such as future income. The data are 
assumed to have a joint distribution
$\Prob(Y,\s,\x_{o},\x_{u})$, from which $n$ observations are drawn.
The fairness assumption requires that $\s$ is not predictive for the response 
given full information:
\[\Prob(Y|\s,\x_{o},\x_{u}) = \Prob(Y|\x_{o},\x_{u}).\]
It is important to posit the existence of both observed
and unobserved legitimate covariates to capture the often observed
relationship between sensitive covariates and the response. Specifically,
observed data often show 
\[\Prob(Y|\s,\x_{o}) \neq \Prob(Y|\x_{o}),\]
which violates the fairness assumption.

These assumptions are captured succinctly for various models using the path 
diagrams of Table \ref{tab:diagrams}. Single headed arrows show direct 
effects while dashed, double-headed arrows indicate correlations which 
are potentially unfair. Observed data are often only representable by a fully 
connected graph which contains no conditional independence properties. This 
observed distribution can be generated 
from multiple full-information models. The first possible representation of the 
full data is an unrestricted model. In this case, sensitive covariates are not 
conditionally independent of the response given full information. The fairness 
assumption is enforced by assuming that sensitive information is conditionally 
independent of the response given full information. Under the fair, 
full-information model, the apparent importance of sensitive information in the 
observed data is only due to unobserved covariates.

One objection to FEO is the assumption that all $\x$ covariates are
legitimate. Thus, while the response may be explained in terms of $\x$ without
recourse to $\s$, that is only because the covariates $\x$ are the result of
structural discrimination. This class of
``potentially illegitimate'' or ``suspect'' covariates is denoted by $\w$ and 
can be used to estimate merit, but only in such a way that does not distinguish 
between groups in $\s$. This treats $ \w$ as proxy variables for missing 
information. The setting for which all legitimate covariates are considered 
suspect is given in the second row of Table \ref{tab:diagrams}. Note again that 
the observed data model can be the result of multiple full-information models. 
The final row of Table \ref{tab:diagrams} shows the fairness assumption in a 
model that includes both legitimate and suspect covariates.

\section{Creating Impartial Estimates}
\label{sec:create}

In this section, we demonstrate one simple way to estimate the models of 
Section 
\ref{sec:models}. We define impartiality with respect to the linear projection 
of $Y$ on a set of covariates $\bf{v}$:
\[Y = {\bf v}'\beta + \epsilon,\]
where the error term $\epsilon$ satisfies
\[\E[\epsilon] = 0, \quad \Cov({\bf v},\epsilon) = \bf{0}.\]
This allows core ideas to be fully explained in a familiar
setting as well as clear decompositions of covariate effects.
While our definition uses linear projections, estimates are not 
required to be linear. Section \ref{sec:data} uses the notion of suspect 
covariates to correct ``black box'' estimates.

For clarity, we introduce the full estimation procedure in stages. Collect the 
observations into matrices $\Y$, $\bS$, $\X$, and $\W$, and consider the 
\emph{estimated} response given various subsets of covariate groups. 
Estimates will be decomposed into various components using projections. For a 
matrix $M$, $\Hm_\mathbf{M}$ is the projection matrix onto the column-span of 
$\mathbf{M}$. We use bracket notation to indicate when two matrices are joined 
column-wise, e.g. $\mathbf{M} = [\X_o,\W]$ contains the columns of both $\X_o$ 
and $\W$. All covariates are assumed to have mean 0 as we include an intercept 
in our model. Covariate matrices of each covariate type are separated
into the portion correlated with the others and the component which is 
orthogonal to them. We will refer to these as the ``shared'' and ``unique'' 
components, respectively. While the decomposition is standard, this is perhaps 
a non-standard presentation. It is important to note that the coefficient for 
each group is computed only from the unique component of that group in the 
model considered. For example, if all covariate groups are used, let $\X_{o,a} 
= (\I - \Hm_{[\bS,\W]})\X_o$ be $\X_o$ ``adjusted'' for the other covariates in 
the model. The least squares estimate of its parameter in this model is given 
by $\hat\beta_{x,t} = (\X_{o,a}'\X_{o,a})^{-1}\X_{o,a}'\Y$, 
provided the inverse exists. In what follows, we put additional subscripts on 
parameters which depend on the model considered. This highlights that 
parameters 
are different depending on the model in which they are estimated. 
Lastly, note that ``hats'' are used to indicate estimated 
values.

First, consider a model with only $\bS$ and $\X$ as predictors 
\eqref{eqn:feo-mod} and the resulting decomposition of the estimates 
\eqref{eqn:decomp-xs}. This is the FEO model as all non-sensitive covariates 
are 
considered legitimate.
\begin{IEEEeqnarray}{rCcCcCc}
	Y & = & \multicol{5}{l}{\beta_{0,f} + \beta_{s,f}^\top\s +
	\beta_{x}^\top\x_{o} + \epsilon_f} \label{eqn:feo-mod}\\
\hat Y_f & = & \hat\beta_{0,f} & + & 
  \underbrace{\Hm_{\X_o} \bS}_{di}\hat\beta_{s,f} + 
  \underbrace{(\I - \Hm_{\X_o})\bS}_{dt}\hat\beta_{s,f}
  &+& \underbrace{\Hm_\bS \X_o}_{sd^+} \hat\beta_{x,f} + 
  \underbrace{(\I - \Hm_\bS)\X_o}_{u}\hat\beta_{x,f} \label{eqn:decomp-xs}
\label{eqn:decomp-s}
\end{IEEEeqnarray}
By decomposing the estimates as in equation (\ref{eqn:decomp-s}), we can 
identify components which are of philosophical and legal interest. The term 
$dt$ 
captures the disparate treatment effect: it is the component of the estimate 
which is due to the unique variability of $\bS$. Given the data generating 
model in Table \ref{tab:diagrams}, we assume that the apparent importance of 
$\bS$ (signified by the magnitude of $\hat \beta_{s,f}$) is due to excluded 
covariates; however, it is identified by $\bS$ in the observed data. While this 
may be a ``sufficiently accurate generalization'' in that the coefficient may 
be statistically 
significant, this is considered to be illegal statistical discrimination 
\cite{RacialDiscrimination}. The term $di$ captures the disparate impact 
effect. We refer to it as the \emph{informative} redlining effect (as it is due 
to an legitimately informative covariate) in order to 
contrast it with an effect identified shortly. Intuitively, it is the misuse of 
informative covariates and is the result of the ability to estimate $\bS$ with 
other covariates. It is important that the adjustment is identified by 
variability in $\bS$ instead of $\X_o$. If one merely ignores $\bS$ completely 
in estimation, the term $di$ will still be included in the final estimate, 
resulting in an estimate which is not impartial.

Previous discussions of redlining do not distinguish between the terms $di$ and 
$sd+$ in equation \eqref{eqn:decomp-xs} \cite{Kam+12b,KamZC13}, because they 
are both due to the correlation between $\X_o$ and $\bS$. It is clear that they 
are different, however, as $\Hm_{\X_o}\bS$ is in the space spanned by $\X_o$ 
and $\Hm_\bS\X_o$ is in the space spanned by $\bS$. Furthermore, the 
coefficients attached to these terms are estimated from different sources. 
Intuition may suggest we remove all components in the space spanned by $\bS$, 
but this is often incorrect. The term $sd+$ cannot be excluded in many settings 
because it accounts for the group means of $\X$. Excluding $sd+$ implies that 
the \emph{level} of $\X$ is not important but that an individual's 
\emph{deviation} from their group mean is. This makes group membership a 
hindrance or advantage and is inappropriate for a legitimate covariate. 
Therefore, in this model $sd+$ must be included.

Second, consider a model with only $\bS$ and $\W$ as predictors 
\eqref{eqn:seo-mod} and the resulting decomposition of the estimates 
\eqref{eqn:decomp-ws}:
\begin{IEEEeqnarray}{rCcCcCc}
  Y & = & \multicol{5}{l}{\beta_{0,p} + \beta_{s,p}^\top\s +
\beta_{w,p}^\top \w + \epsilon_p,} \label{eqn:seo-mod}\\
 \hat Y_p & = & \hat\beta_{0,p} & + & 
  \underbrace{\Hm_{\W}\bS}_{di}\hat\beta_{s,p} + 
  \underbrace{(\I - \Hm_{\W})\bS}_{dt}\hat\beta_{s,p}
  &+& \underbrace{\Hm_\bS \W}_{sd^-} \hat\beta_{w,p} + 
  \underbrace{(\I - \Hm_\bS)\W}_{u}\hat\beta_{w,p}.\label{eqn:decomp-ws}
\end{IEEEeqnarray}
In equation \eqref{eqn:decomp-ws}, $sd-$ addresses the concern that group 
differences in covariate $\W$ are potentially discriminatory, as contrasted 
with $sd+$ in \eqref{eqn:decomp-xs}. For example, if $\W$ is location, $sd-$ 
measures racial differences between neighborhoods. Given that proxy 
variables $\W$ are not considered directly informative, it is unclear what 
these differences can legitimately contribute. If there is racial bias in 
neighborhood demographics, using this information would perpetuate this 
discrimination. Ensuring that this does not occur requires removing $sd-$ from 
the estimates of $\Y$. This identifies a new type of redlining effect that we 
call \emph{uninformative} redlining (as it is due to a proxy variable); it is 
the sum of $di$ and $sd-$. Uninformative redlining can be identified visually 
using the graphs in Table \ref{tab:diagrams}. Fairness constrains the 
information contained in the arrow $\s \to Y$ as well as information conveyed 
in 
the path $\s \to \w \to Y$. This 
is because $\s \to \w$ is potentially discriminatory. Therefore, impartial 
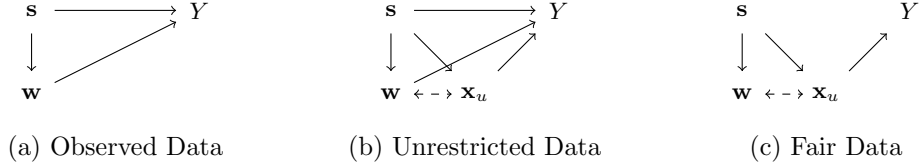
estimates with suspect or proxy variables only use the unique variability in 
$\W$. An important consequence of this SEO model is that average estimates are 
the same for different groups of $\s$. This is an alternate construction of the 
initial estimates used by \cite{Cald+13}.

Lastly, the final model includes all covariate groups $\bS$, $\X$, and $\W$ as 
predictors \eqref{eqn:total-mod} and the resulting estimates 
\eqref{eqn:decomp-xws}:
\begin{IEEEeqnarray*}{rCcCcCc}
	Y & = & \multicol{5}{l}{\beta_{0,t} + \beta_{s,t}^\top\s +
	\beta_{x,t}^\top\x_{o} + \beta_{w,t}^\top\w + \epsilon_t.} 
	\yesnumber\label{eqn:total-mod}\\
	\hat \Y & = & \hat\beta_{0,t}  & + & 
	\underbrace{\Hm_{[\X_o,\W]} \bS}_{di}\hat\beta_{s,t} &+&
	\underbrace{(\I - \Hm_{[\X_o,\W]})\bS}_{dt}\hat\beta_{s,t} \\
	&&&+&\underbrace{\Hm_{[\bS,\W]} \X_o}_{sd^+}\hat\beta_{x,t}
	&+&\underbrace{(\I - \Hm_{[\bS,\W]})\X_o}_{u}\hat\beta_{x,t}\\
	&&&+&\underbrace{\Hm_{[\X_o,\bS]}\W}_{sd^-,sd^+}\hat\beta_{w,t} &+&
	\underbrace{(\I-\Hm_{[\X_o,\bS]})\W}_{u}\hat\beta_{w,t}
	\yesnumber\label{eqn:decomp-xws}
\end{IEEEeqnarray*}

While the majority of the terms in the above display mirror the previous 
discussion, the suspect covariates display a more complex behavior. The unique 
component can still be considered additional information orthogonal to the 
sensitive covariates, but the component correlated with other covariates is 
labeled both $sd+$ and $sd-$ to indicate that this combines both legal and 
illegal forms of statistical discrimination. The notation $\Hm_{[\X_o,\bS]}\W$ 
indicates that the shared component is the best linear estimate of $\W$ given 
both $\bS$ and $\X$. As $\W$ is a suspect covariate, we can remove $sd-$ while 
retaining $sd+$ by producing an impartial estimate of $\W$. In this case, $\W$ 
has taken the place of $\Y$ as the response in an FEO model that only contains 
$\bS$ and $\X_o$.

Removing the components labeled $di$, $dt$, and $sd-$ can be accomplished 
through the following procedure.
\begin{defn}[Impartial Estimate]
	\label{defn:impartial-reg}
	Linearly impartial estimates are created with the following multi-step 
procedure, where ``estimate" means ``estimate via least-squares:"
	\begin{enumerate}
		\item Estimate the model (\ref{eqn:total-mod}) to produce
		$\hat\beta_{0}$, $\hat\beta_{s}$, $\hat\beta_{x}$, and
		$\hat\beta_{w}.$
		\item Create an impartial estimate of each element of $\w$.
		     \begin{itemize}
		      \item[a.] Estimate $\w = \lambda_0 +\Lambda_s^\top\s
		 +\Lambda_{x}^\top\x_o +\eta$.
		      \item[b.] Set $\hat\w = \hat\lambda_0 + \hat\Lambda_{x}^\top\x_o$.
		      \item[c.] Collect the estimates, $\hat\w$, as $\hat\W$.
		     \end{itemize}
		\item Set $\hat Y = \hat\beta_{0} + \X_o\hat\beta_{x} +
		\hat\W\hat\beta_{w} + (\I-\Hm_{[\X_o,\bS]})\W\hat\beta_{w}$.
	\end{enumerate}
\end{defn}

\section{Simple Example: FEO vs SEO}
\label{sec:simple-ex}

This section provides a simplified example to compare the estimates implied by 
FEO and SEO. Without a proper data-generation narrative, ``fair'' estimates can 
appear decidedly \emph{unfair}. Consider a simple example of credit score 
modeling that has only two covariates: education level, $\x$, and sensitive 
group, $\s$. Suppose the data is collected on individuals who took out a loan of 
a given size, that higher education is indicative of better repayment, and that 
education is split into two categories: high and low. To see the relevant 
issues, $\s$ and $\x$ need to be associated. The two sensitive groups will be 
written as $\s_+$ and $\s_-$ to indicate which group, on average, has higher 
education: the majority of $s_-$ have low education and the majority of $s_+$ 
have high education. The response is the indicator of default, $D$.

\begin{table*}[t]
\centering 
  \caption{Simplified loan repayment data and estimated default probabilities.}
  \label{tab:simple-data}
  \begin{tabular}{|l|ll|ll|c|c|}
    \hline Education ($\Prob(x)$) & \multicolumn{2}{c|}{Low (.6)} &
    \multicolumn{2}{c|}{High (.4)}&& \\
    Group ($\Prob(s|x)$) & $s_-$ (.75) & $s_+$ & $s_-$ (.25) & $s_+$ &
    Total&\\\hline 
    Default Yes & 225 & 60 & 20 & 30 & 335&\\
    Default No & 225 & 90 & 80 & 270 & 665&\\\hline
    \multicolumn{6}{c}{}\\
    \hline&\multicolumn{4}{c|}{$\hat\Prob(\text{Default 
YES}|x_i,s_i)$}&DS&RMSE\\\hline
    Full Model& .5 & .4 & .2 & .1& -.25 & 13.84\\
    Exclude $\s$ & .475 & .475 & .125 & .125 & -.17 & 13.91\\
    FEO & .455 & .455 & .155 & .155& -.15 & 13.93\\
    SEO & .39 & .535 & .09 & .235& 0.00 & 14.37\\
    Marginal & .35 & .35 & .35 & .35 & 0.00 & 14.93\\\hline
  \end{tabular}
\end{table*}

Synthetic data and estimates are provided in Table \ref{tab:simple-data}, in 
which there exist direct effects for both $\s$ and $\x$. This is consistent 
with 
the observed data graphs in Table \ref{tab:diagrams}. Five possible estimates 
are compared in Table \ref{tab:simple-data}: the full OLS model, the restricted 
regression which excludes $\s$, the FEO model in which education is considered 
a 
legitimate covariate, the SEO model in which education is considered a suspect 
covariate, and the marginal model which estimates the marginal probability of 
default without any covariates. Estimates are presented along with the 
in-sample 
RMSE from estimating the true default indicator and the discrimination score 
(DS) \cite{CalV10}.
While we have argued that ``discrimination score'' is at times a misnomer since 
it does not separate explainable from discriminatory variation, it provides a 
useful perspective given its widespread use in the literature. Furthermore, as 
one often discusses discrimination toward groups, it is nevertheless helpful to 
gauge the difference between estimates between groups.

Since education is the only covariate that can measure similarity, one would 
expect that estimates should be constant for individuals with the same 
education. This is easily accomplished by the legal prescription of input 
scrutiny by excluding $\s$. If the information is not observed, it cannot lead 
to disparate treatment 
directly related to group membership. The FEO model satisfies this as well. As 
seen in equation \eqref{eqn:decomp-xs}, the only 
difference between the two estimates is the coefficient on $\x$. Said 
differently, the term $di$ in equation (\ref{eqn:decomp-s}) lies in the space 
spanned by $\x$. Therefore its removal only changes estimates for education 
groups. Excluding $\s$ permits redlining because it increases the estimated 
disparity between low- and high-education groups. This disproportionately 
effects those in $\s_-$ as they constitute the majority of the low education 
group. The FEO estimates result in some average differences between groups, but 
this is acceptable if the association between $\x$ and $\s$ is benign. This 
accurately measures the proportional differences desired by \cite{Banks01} for 
fair treatment.

The SEO estimates appear counter-intuitive: although $\s_+$ performs better
in our data set even after accounting for education, these estimates
predict the opposite. Understanding this requires accepting the world view
implicit in the SEO estimates: average education differences between groups are
the result of structural discrimination. Members of $\s_-$ in the high
education group have a much higher education than average for $\s_-$. Similarly,
members of $\s_+$ who are in the high education group have a higher  education
than average for $\s_+$, but not by as much. The magnitude of these differences
is given importance, not the education level. 

As an example where these deviations are given importance, consider the
college admissions process in the United States and the two explanatory 
covariates ``class rank'' and ``SAT score.'' The SAT is a standardized test 
commonly used in the United States to assess students' readiness for 
university, and it is under scrutiny for doing more to measure disparities 
in students’ learning opportunities, e.g. wealth and race, than 
college preparedness \cite{JencksP98,Geiser17}. The SAT score provides 
information on where an applicant lies in the national test score distribution, 
whereas class rank specifies their location in the local grade distribution. 
Considering class rank is equivalent to placing importance on the deviation 
between a student and others much more likely to be in a comparable situation. 
Similarly, the SAT score could be used to only measure differences 
\emph{within} a group. Conceptually, this is what was done in the ``Strivers'' 
proposal, wherein students were termed ``strivers'' when they 
significantly outperformed the expected score based on socioeconomic and 
structural factors \cite{CarneH98}.


In our example, the SEO model balances the differences in education 
distributions, resulting in both groups having the same average estimated 
default. This is seen in the discrimination score of 0. Without claiming that 
education is partially the result of structural differences, the SEO estimates 
discriminate against $\s_+$. Other methods to achieve group fairness 
produce estimates relevantly similar to SEO in this regard. Furthermore, if the 
structural effects are not such that all $\s_+$ individuals are given a benefit 
or not all $\s_-$ individuals receive a detriment, then these models are merely 
approximations of the fair correction. An ideal protected or sensitive 
covariate 
$\s$ is exactly that which accounts for differences in the opportunity of being 
high merit. This is in line with Rawls' conception of equality of 
opportunity \cite{Rawls01}.

The SEO estimates show another important property: their RMSE is lower than
that of the marginal estimate of default while still minimizing the
discrimination score. Therefore, if a bank is required to minimize differences
between groups in the interest of fairness, it would rather use the SEO
estimates than the marginal estimate. SEO still acknowledges that education is 
an informative predictor and contains an education effect. Furthermore, 
equality 
of opportunity is not satisfied when marginal estimates are used because all 
merit information is ignored. See \cite{sep-eo} for a more detailed discussion.

\section{Data Illustrations with Black-Box Estimates}
\label{sec:data}

This section present results on two criminology data sets. For a detailed 
discussion of the trade-offs between different fairness measures 
in criminology see \cite{Berk+21}. 
We show the performance of various 
procedures using the discrimination score (average predicted difference between 
groups), the root mean squared prediction error, as well as the positive and 
negative residual differences (PRD and NRD, respectively):
\begin{IEEEeqnarray*}{rCl}
  PRD & := & \left| \frac{1}{n_1} \sum_{i\in S_1} \max\{0, Y_i-\hat Y_i\} -
  \frac{1}{n_0} \sum_{i\in S_0} \max\{0, Y_i-\hat Y_i\} \right|\\
  NRD & := & \left| \frac{1}{n_1} \sum_{i\in S_1} \min\{0, Y_i-\hat Y_i\} -
  \frac{1}{n_0} \sum_{i\in S_0} \min\{0, Y_i-\hat Y_i\} \right|.
\end{IEEEeqnarray*}
These measures play the role of false positive and false negative rates for 
regression problems \cite{Cald+13}. All statistics shown in the following 
subsections are computed out-of-sample using 5-fold cross-validation and 
averaged over 12 repetitions.

Lastly, to correct a ``black-box'' estimate, suppose that we have an estimate 
of the response, $Y^\dag$, given by an unknown model with unknown inputs. The 
model may use sensitive information to be intentionally or unintentionally 
discriminatory. While potentially informative, there 
is no guarantee that the estimates are impartial. This identically matches the 
description of suspect covariates. Therefore, if we treat $Y^\dag$ as a suspect 
covariate, its information can be used but not in a way that makes distinctions 
between protected groups. For simplicity, we only include estimates from a 
support vector regression \cite{Awad2015} or random forest algorithm 
\cite{Brei01} as they are high-performing, off-the-shelf ``black boxes.'' 
Furthermore, they are allowed to use $\s$ and $\w$ in order to demonstrate that 
estimates can be easily corrected.

We compare our models with the propensity score stratification methods of 
\cite{Cald+13} as well as  the convex optimization approaches of 
\cite{Heidari+18,Heidari+19}. \cite{Heidari+18} is motivated by social 
welfare considerations, places an upper bound on prediction error, and controls 
the PRD and NRD. \cite{Heidari+19} requires a prespecified utility function to 
be defined for all groups and explicitly tries to maximize the utility for both 
groups while placing an upper bound on prediction error. These three methods 
will be referred to as $propensity$, $welfare$, and $utility$, due to the 
motivations for the frameworks. These methods are compared to 
impartial estimates that use various covariate specifications for $\s$, $\x$, 
and $\w$. The $exclude$ model merely excludes $\s$ during estimation, the $FEO$ 
model treats all non-sensitive features as legitimate, the 
$SEO$ model treats all non-sensitive features as suspect, and the $mixed$ model 
treats some as legitimate and some as suspect. We separately consider each 
model when they are given additional black-box estimates
which are treated as a suspect covariate ($\_rf$ or $\_svr$).

In all data cases, $\Y$ is rescaled to lie in $[0,1]$ and larger values 
correspond to better estimates, i.e. lower 
estimates of crime and fewer days jailed. This is required to use the utility 
function specified in \cite{Heidari+19}. As acknowledged therein, the choice of 
this function has large impact on estimates as seen below. The final example 
uses the same data set used in \cite{Heidari+19} to promote easier comparison. 
Rescaling in this way yields small baseline values for the mean difference in 
$\Y$ as well as the standard deviation of $\Y$. As such, small reductions in 
prediction error can still be sizable on a more natural scale.

The first data set contains information from parolees. The guiding question is 
whether men are more likely to do hard time holding constant age, the 
neighborhood in which the live, and prior record. In particular, the goal is to 
provide estimates of number of days in jail which are impartial with respect to 
sex. This data set contains information from approximately 83,000 individuals on 
probation after excluding observations that with zero previous jail days as well 
as minors. Covariate information includes age, sex, and prior record information 
such as the number of violent priors as an adult or juvenile. We also consider 
adding a random forest estimate that is not constrained in its use of covariate 
information. Lastly, as all covariates and the response exhibit long right 
tails, all variables are log transformed. The only covariate treated as suspect 
is age, as beyond attributing to prior record, it is unclear what direct 
information this could contain.

For this data set, the standard deviation of the transformed response is 0.227 
with a mean difference of 0.05 between groups. Therefore, in general, men have 
served longer sentences without conditioning on prior record. Table 
\ref{tab:jail} shows that the random forest estimates do not appreciably improve 
prediction in this setting. That being said, the models already perform as well 
as all competitors in terms of prediction error. While the differences are 
slight, we can see that driving the DS to zero again results in small increases 
in the PRD and NRD.

\begin{table}[ht]
\centering
\caption{Impartial Estimates of Jail Days}
\label{tab:jail}
\begin{tabular}{lrrrrr}
  \toprule
 Model & DS & PRD & NRD & RMSE\\ 
  \midrule
  exclude & 0.02 & 0.02 & 0.01 & 0.14 \\ 
  FEO & 0.02 & 0.02 & 0.01 & 0.14 \\ 
  SEO & {\bf 0.00} & 0.03 & 0.02 & 0.14 \\ 
  mixed & 0.02 & 0.02 & 0.01 & 0.14 \\ 
  \midrule
  FEO\_rf & 0.02 & 0.02 & 0.01 & 0.13 \\ 
  SEO\_rf & {\bf 0.00} & 0.03 & 0.02 & 0.13 \\ 
  mixed\_rf & 0.02 & 0.02 & 0.01 & 0.13 \\ 
  \midrule
  propensity & 0.02 & 0.01 & 0.02 & 0.22 \\ 
  utility & 0.07 & 0.01 & 0.01 & 0.17 \\ 
  welfare & 0.02 & 0.00 & 0.02 & 0.14 \\ 
  \bottomrule
\end{tabular}
\end{table}

Our second illustration uses the Crime and Communities data set also considered 
in \cite{Heidari+19}. This allows for a more direct comparison as the authors 
specify a utility function in this case, removing the largest open input to 
using the method. The data set contains information on 1,994 communities such 
as demographic statistics (race, immigrant, and age distributions), law 
enforcement (budget, racial distribution of officers, etc), economic (income, 
unemployment, home-ownership, etc). In total, there are 100 predictive 
covariates which can be used to estimate the per capita number of violent 
crimes. Impartial estimates of crime in this case directly links back to the 
original redlining example presented in the introduction: lower estimates of 
crime provide more incentive for investment, etc.

Similar to \cite{Heidari+19}, we label a community as a member of the protected 
group if more than 25\% of the residents are black. Specifying the groups this 
way as opposed to merely non-Caucasian highlights an important concept behind 
the current work: it is possible to predict the proportion of black residents 
in 
a community using the other covariates. In fact, there are 14-20 covariates 
which can be used either separately or together to reduce the error sum of 
squares for predicting the proportion of black residents by 95\%. As such, this 
data set provides our first sincere case with suspect covariates which fall 
into 
various categories. Examples of predictive economic indicators include the 
percentage of households with income from investments, social security, or from 
public assistance. Others include percentage of people born in the same state, 
or living in the same house as 5 years before, as well as number of people in 
shelters or on the street. The conceptual difficulty is that there may be some 
debate about whether these covariates are legitimate. For comparison, we 
consider all such covariates as suspect in the mixed model in Table 
\ref{tab:crime}.

\begin{table}[ht]
\centering
\caption{Impartial Estimates of Crime}
\label{tab:crime}
\begin{tabular}{rlrrrrr}
\toprule
 Model & DS & PRD & NRD & RMSE\\ 
\midrule
  exclude & 0.30 & 0.04 & 0.04 & 0.14 \\ 
  FEO & 0.28 & 0.02 & 0.05 & 0.14 \\ 
  SEO & {\bf 0.00} & 0.08 & 0.23 & 0.19 \\ 
  mixed & 0.23 & 0.00 & 0.08 & 0.14 \\ 
  \midrule
  FEO\_svr & 0.28 & 0.01 & 0.04 & 0.11 \\ 
  SEO\_svr & 0.00 & 0.08 & 0.23 & 0.17 \\ 
  mixed\_svr & 0.23 & 0.01 & 0.07 & 0.11 \\ 
  \midrule
  propensity & 0.27 & 0.02 & 0.06 & 0.15 \\ 
  utility & 0.19 & 0.07 & 0.05 & 0.19 \\ 
  welfare & 0.30 & 0.01 & 0.02 & 0.14 \\
\bottomrule
\end{tabular}
\end{table}

We now see clear differences in performance of various impartial models. 
Importantly, the predictive performance of the mixed model matches that of the 
FEO model while reducing the mean difference between groups. Furthermore, there 
is also now a penalty in prediction for switching to a full SEO model which 
considers no legitimate covariates. As the mean difference in the transformed 
response is 0.31, it is clear that many models make little to no improvement in 
this regard. On the other hand, including the black box estimates improves 
predictive performance dramatically. While the welfare model does achieve 
good predictive performance, the mean difference is hardly changed. As a final 
note, we observe that the propensity model performs better, as it is no longer 
able to perfectly classify the protected groups based on the remaining 
legitimate covariates. That being said, if one wants the mean difference 
between 
groups to be small, the SEO models achieve this exactly.

\section{Discussion}
\label{sec:disc}

This paper provides a clear statistical theory of impartial estimation and 
explainable variability through the analysis of proxy variables. Covariate 
groups can only be specified for use in FADM if one has a clear understanding 
of 
the implications of the decisions. By considering full-data scenarios in which 
our fairness assumption is satisfied, we concretely describe the role of proxy 
variables and their allowable use in FADM. This also provides connections 
to legal concepts such as our newly identified ``uninformative redlining'' 
effect and distinctions between various types of statistical discrimination.

\bibliography{/home/johnson/Dropbox/Research/Bib_Stuff/Bibliography}

\begin{thebibliography}{10}

\bibitem{Adler18}
P.~Adler, C.~Falk, S.~A. Friedler, T.~Nix, G.~Rybeck, C.~Scheidegger, B.~Smith,
  and S.~Venkatasubramanian.
\newblock Auditing black-box models for indirect influence.
\newblock {\em Knowl. Inf. Syst.}, 54(1):95--122, 2018.

\bibitem{sep-eo}
R.~Arneson.
\newblock Equality of opportunity.
\newblock In E.~N. Zalta, editor, {\em The Stanford Encyclopedia of
  Philosophy}. Summer 2015 edition, 2015.

\bibitem{Awad2015}
M.~Awad and R.~Khanna.
\newblock {\em Support Vector Regression}, pages 67--80.
\newblock Apress, Berkeley, CA, 2015.

\bibitem{Banks01}
R.~R. Banks.
\newblock Race-based suspect selection and colorblind equal protection doctrine
  and discourse.
\newblock {\em UCLA Law Review}, 48, 2001.

\bibitem{BenH19}
S.~Benthall and B.~D. Haynes.
\newblock Racial categories in machine learning.
\newblock In {\em Proceedings of the conference on fairness, accountability,
  and transparency}, pages 289--298, 2019.

\bibitem{Berk+21}
R.~Berk, H.~Heidari, S.~Jabbari, M.~Kearns, and A.~Roth.
\newblock Fairness in criminal justice risk assessments: The state of the art.
\newblock {\em Sociological Methods \& Research}, 50(1):3--44, 2021.

\bibitem{RacialDiscrimination}
R.~M. Blank, M.~Dabady, and C.~F. Citro, editors.
\newblock {\em Measuring Racial Discrimination}.
\newblock National Research Council, 2004.

\bibitem{Brei01}
L.~Breiman.
\newblock Random forests.
\newblock {\em Machine Learning}, 45(1):5--32, 2001.

\bibitem{Brock06}
J.~Brockner.
\newblock It's so hard to be fair.
\newblock {\em Harvard business review}, 84(3):122, 2006.

\bibitem{Cald+13}
T.~Calders, A.~Karim, F.~Kamiran, W.~Ali, and X.~Zhang.
\newblock Controlling attribute effect in linear regression.
\newblock In {\em Data Mining (ICDM), 2013 IEEE 13th International Conference
  on}, pages 71--80, Dec 2013.

\bibitem{CalV10}
T.~Calders and S.~Verwer.
\newblock Three naive bayes approaches for discrimination-free classification.
\newblock {\em Data Mining and Knowledge Discovery}, 21(2):277--292, 2010.

\bibitem{CarneH98}
A.~P. Carnevale and E.~Haghighat.
\newblock Selecting the strivers: a report on the preliminary results of the
  ets ``educational strivers'' study.
\newblock {\em Hopwood, Bakke, and beyond: Diversity on our nation{\rq}s
  campuses}, pages 122--128, 1998.

\bibitem{Chiappa2018}
S.~Chiappa and W.~S. Isaac.
\newblock A causal bayesian networks viewpoint on fairness.
\newblock In {\em IFIP International Summer School on Privacy and Identity
  Management}, pages 3--20. Springer, 2018.

\bibitem{Eubanks18}
V.~Eubanks.
\newblock {\em Automating inequality: How high-tech tools profile, police, and
  punish the poor}.
\newblock St. Martin's Press, 2018.

\bibitem{Geiser17}
S.~Geiser.
\newblock Norm-referenced tests and race-blind admissions: the case for
  eliminating the sat and act at the university of california.
\newblock {\em Center for Studies in Higher Education. Research and Occasional
  Paper Series (ROPS). CSHE}, 15, 2017.

\bibitem{Gillis21}
T.~B. Gillis.
\newblock The input fallacy.
\newblock {\em Minnesota Law Review}, 2022.
\newblock Forthcoming.

\bibitem{Grgic+16}
N.~Grgic-Hlaca, M.~B. Zafar, K.~P. Gummadi, and A.~Weller.
\newblock The case for process fairness in learning: Feature selection for fair
  decision making.
\newblock In {\em NIPS symposium on machine learning and the law}, volume~1,
  page~2, 2016.

\bibitem{HajD13}
S.~Hajian and J.~Domingo-Ferrer.
\newblock A methodology for direct and indirect discrimination prevention in
  data mining.
\newblock {\em IEEE Trans. Knowl. Data Eng.}, 25(7):1445--1459, 2013.

\bibitem{Heidari+18}
H.~Heidari, C.~Ferrari, K.~Gummadi, and A.~Krause.
\newblock Fairness behind a veil of ignorance: A welfare analysis for automated
  decision making.
\newblock In S.~Bengio, H.~Wallach, H.~Larochelle, K.~Grauman, N.~Cesa-Bianchi,
  and R.~Garnett, editors, {\em Advances in Neural Information Processing
  Systems}, volume~31. Curran Associates, Inc., 2018.

\bibitem{Heidari+19}
H.~Heidari, M.~Loi, K.~P. Gummadi, and A.~Krause.
\newblock A moral framework for understanding fair ml through economic models
  of equality of opportunity.
\newblock In {\em Proceedings of the Conference on Fairness, Accountability,
  and Transparency}, FAT* '19, pages 181--190, New York, NY, USA, 2019.
  Association for Computing Machinery.

\bibitem{Holl03}
P.~W. Holland.
\newblock Causation and race.
\newblock {\em ETS Research Report Series}, 2003(1):i--21, 2003.

\bibitem{HurleyA16}
M.~Hurley and J.~Adebayo.
\newblock Credit scoring in the era of big data.
\newblock {\em Yale JL \& Tech.}, 18:148, 2016.

\bibitem{JencksP98}
C.~Jencks and M.~Phillips.
\newblock The black-white test score gap: An introduction.
\newblock {\em The Black-White test score gap}, 1(9):26, 1998.

\bibitem{KamCP10}
F.~Kamiran, T.~Calders, and M.~Pechenizkiy.
\newblock Discrimination aware decision tree learning.
\newblock In {\em Proceedings of the 2010 IEEE International Conference on Data
  Mining}, ICDM '10, pages 869--874, Washington, DC, USA, 2010. IEEE Computer
  Society.

\bibitem{KamZC13}
F.~Kamiran, I.~Zliobaite, and T.~Calders.
\newblock Quantifying explainable discrimination and removing illegal
  discrimination in automated decision making.
\newblock {\em Knowl. Inf. Syst.}, 35(3):613--644, 2013.

\bibitem{Kam+12b}
T.~Kamishima, S.~Akaho, H.~Asoh, and J.~Sakuma.
\newblock Fairness-aware classifier with prejudice remover regularizer.
\newblock In P.~A. Flach, T.~D. Bie, and N.~Cristianini, editors, {\em
  ECML/PKDD (2)}, volume 7524 of {\em Lecture Notes in Computer Science}, pages
  35--50. Springer, 2012.

\bibitem{Kilbertus17}
N.~Kilbertus, M.~{Rojas Carulla}, G.~Parascandolo, M.~Hardt, D.~Janzing, and
  B.~Sch{\"o}lkopf.
\newblock Avoiding discrimination through causal reasoning.
\newblock In I.~Guyon, U.~V. Luxburg, S.~Bengio, H.~Wallach, R.~Fergus,
  S.~Vishwanathan, and R.~Garnett, editors, {\em Advances in Neural Information
  Processing Systems}, volume~30. Curran Associates, Inc., 2017.

\bibitem{KozJL21}
N.~Kozodoi, J.~Jacob, and S.~Lessmann.
\newblock Fairness in credit scoring: Assessment, implementation and profit
  implications.
\newblock {\em European Journal of Operational Research}, 2021.

\bibitem{Kusner17}
M.~J. Kusner, J.~Loftus, C.~Russell, and R.~Silva.
\newblock Counterfactual fairness.
\newblock In I.~Guyon, U.~V. Luxburg, S.~Bengio, H.~Wallach, R.~Fergus,
  S.~Vishwanathan, and R.~Garnett, editors, {\em Advances in Neural Information
  Processing Systems}, volume~30. Curran Associates, Inc., 2017.

\bibitem{morgan_winship_14}
S.~L. Morgan and C.~Winship.
\newblock {\em Counterfactuals and Causal Inference: Methods and Principles for
  Social Research}.
\newblock Analytical Methods for Social Research. Cambridge University Press, 2
  edition, 2014.

\bibitem{Noble18}
S.~U. Noble.
\newblock {\em Algorithms of oppression}.
\newblock New York University Press, 2018.

\bibitem{Pearl09}
J.~Pearl.
\newblock {\em Causality: Models, Reasoning and Inference}.
\newblock Cambridge University Press, 2 edition, sep 2009.

\bibitem{PedRT08}
D.~Pedreschi, S.~Ruggieri, and F.~Turini.
\newblock Discrimination-aware data mining.
\newblock In Y.~B.~L. Li and S.~Sarawagi, editors, {\em KDD}, pages 560--568.
  ACM, 2008.

\bibitem{PopeS11}
D.~G. Pope and J.~R. Sydnor.
\newblock Implementing anti-discrimination policies in statistical profiling
  models.
\newblock {\em American Economic Journal: Economic Policy}, 3(3):206--31,
  August 2011.

\bibitem{R_2019}
{R Core Team}.
\newblock {\em R: A Language and Environment for Statistical Computing}.
\newblock R Foundation for Statistical Computing, Vienna, Austria, 2019.

\bibitem{Ravi+20}
P.~Ravishankar, P.~Malviya, and B.~Ravindran.
\newblock A causal linear model to quantify edge flow and edge unfairness for
  unfairedge prioritization and discrimination removal.
\newblock {\em arXiv preprint arXiv:2007.05516}, 2020.

\bibitem{Rawls01}
J.~Rawls.
\newblock {\em Justice as fairness: A restatement}.
\newblock Harvard University Press, 2001.

\bibitem{sep-rawls}
L.~Wenar.
\newblock John rawls.
\newblock In E.~N. Zalta, editor, {\em The Stanford Encyclopedia of
  Philosophy}. Winter 2013 edition, 2013.

\bibitem{Zem+13}
R.~S. Zemel, Y.~Wu, K.~Swersky, T.~Pitassi, and C.~Dwork.
\newblock Learning fair representations.
\newblock In {\em ICML (3)}, volume~28 of {\em JMLR Workshop and Conference
  Proceedings}, pages 325--333. JMLR.org, 2013.

\end{thebibliography}

\end{document}